# A Trust Model Based Analysis of Social Networks

Surya Nepal, Cécile Paris, Sanat Kumar Bista
CSIRO ICT Centre, Australia
Author Email: firstname.lastname@csiro.au

Wanita Sherchan
IBM Research, Australia
Author Email: wanitash@au.ibm.com

**Abstract**: In this paper, we analyse the sustainability of social networks using *STrust,* our social trust model. The novelty of the model is that it introduces the concept of engagement trust and combines it with the popularity trust to derive the social trust of the community as well as of individual members in the community. This enables the recommender system to use these different types of trust to recommend different things to the community, and identify (and recommend) different roles. For example, it recommends *mentors* using the engagement trust and *leaders* using the popularity trust. We then show the utility of the model by analysing data from two types of social networks. We also study the sustainability of a community through our social trust model. We observe that a 5% drop in highly trusted members causes more than a 50% drop in social capital that, in turn, raises the question of sustainability of the community. We report our analysis and its results.

## 1. Introduction

Social networking sites provide a space in which people can share information, opinions, experiences, interests, can offer each other support, and, more generally, can connect with each other. When Web-based social networks first appeared, they were mostly exploited by individual users to keep in touch with their friends and families (Mika, 2007). With their popularity and phenomenal growth, governments and commercial enterprises have also looked at exploiting their potential to deliver and improve services (Jaeger et al., 2007) (Zappen et al., 2008) (Borchorst et al., 2011). However, not all social networks are successful: indeed, many social networks disappear because they fail to attract enough members or because there are not enough interactions amongst members to retain people. This raises two important questions: (a) what level of interactions ensures the sustainability of a social network, and (b) how to increase the number of members and interactions between them. In this paper, we partly answer these questions through the concepts of *social trust* and *social capital*.

There have been reports in media of many incidents of breaching privacy of individuals through social networks (Gross and Acquisti, 2005) (Dwyer et al., 2007) (Young and Quan-Haase, 2009). Privacy is a very important consideration for the users, and it has a direct impact on the number of members and their interactions. In order to balance the open nature of the social networks and safeguard the privacy concerns of the users, it is important to build *trust communities*. A trust community is a community that creates an environment in which its members can share their thoughts, opinions and experiences in an open and honest way without concerns about their privacy and fear of being judged. We contend that social trust provides an ideal foundation for building trust communities, and thus, ensuring social trust in a network is one way to attract members and increase the number of interactions. Using the concept of social capital, we have developed a social trust model with the aim of building trust communities (Nepal et al., 2012). Social capital refers to the richness of the interactions amongst members, or the interactions from which members derive benefits (Putnam, 1995, Putnam, 2000). In our trust model, we consider two aspects of trust: *Popularity Trust* (PopTrust) and *Engagement Trust* (EngTrust). Popularity trust (Caverlee et al., 2010) refers to the acceptance and approval of a member by others in the community, while engagement trust captures the involvement of someone in the community. We can consider popularity trust to reflect the trustworthiness of a member in the community, and engagement trust how much a member trusts other members in the community. Our model separates these trust values as they can be used to recommend different things and identify different roles in the community. For example, the popularity trust can be used to identify leaders in the community, while recommendation to be friends and mentors can be made using the engagement trust.

In this paper, we build on our previous work, exploring the utility of our model by applying it to real datasets representing Facebook-like social networks. We first describe how the separation of engagement trust and popularity trust enables a system to capture different types of interactions, recommend different things to different members in the community, and identify different roles in the community. Then we use the real datasets to demonstrate that there are indeed people with different roles in a community, so that it is useful to be able to distinguish these roles. Finally, we address the question of sustainability by removing highly trusted members to study the impact on the community. This can help community developers select a target user group (highly engaged and popular) that need to be retained in the community.

The rest of the article is structured as follows. Section 2 describes our framework for building trust communities. Section 3 briefly describes STrust, our proposed social trust model. In Section 4, we introduce a recommender system based on this model. In Section 5, we present our study of the model using real datasets and report our results with respect to the sustainability of the community. We discuss related work in Section 6, and, finally, Section 7 presents some concluding remarks and our future work

## 2. A Framework for Building Trust Communities

In (Nepal et al., 2012), we proposed a framework to build trust communities from social networks. Our framework consists of the following four steps, as shown in Figure 1.

**Personal Background:** A user first registers in a social network and creates an account. The user provides personal details that he or she would like to share with other members in the community, such as email address, date of birth, hobbies, etc. This step also involves choosing an identity, for example, the selection of an avatar, of events of interest, and of preferences for friends. We refer the first step as setting up a *user model*.

**Social Capital:** Here, users build their social capital by interacting with others, for example by befriending someone, providing comments on content put by others, providing a comment on a comment, etc. The purpose of this step is to create an environment for interactions.

**Social Trust:** In this step, the social trust of an individual member and of the community as whole is evaluated based on social capital. We refer to the latter two steps as creating a *social model*.

**Recommendation**: The last step in our approach is the generation of recommendations based on trust. The aim of the recommender is to make the online community relevant to the members so that we can increase the social capital and social trust, which in turn is used again by the recommendation system to recommend new activities or content. This cycle continues, first to build the trust community and then to ensure its sustainability. This is required because trust decays with time (which is sometimes refers to as *aging*) (Wishart et al., 2005 ).

## 3. Social Trust Model (STrust)

In this section, we first introduce the concepts of social trust and social capital. We then define STrust, our social trust model for social networks.

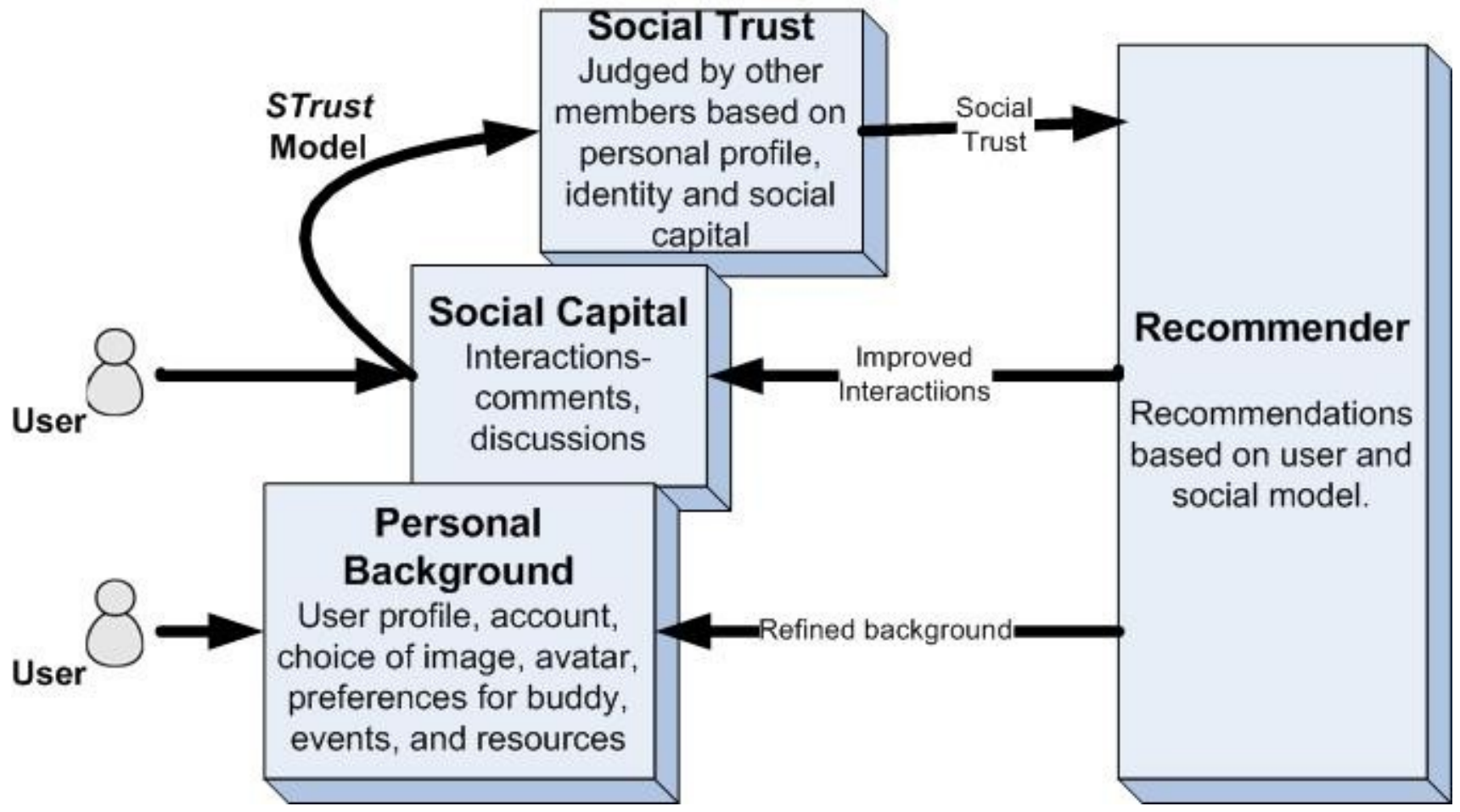

Figure 1. Our framework for building trust communities in social networks

### *3.1 Social Trust*

Following (Singh and Bawa, 2007), we define social trust as the *firm belief in the competence of an entity to act as expected, such that this firm belief is not a fixed value associated with the entity, but rather it is subject to entity's behaviour and applies only within a specific context at any given time*. There are three important aspects of trust. We explain each of these briefly before defining our social trust model.

**User behaviour**: Social trust depends on the behaviour of an individual. In the context of an online community, the behaviour of an individual is derived from his or her interactions in the community. There are two types of interactions: active and passive. Examples of active interactions include having a large number of friends, posting information regularly, replying to other members' posts, etc. Passive interactions include reading posts, reading articles, regular visits to the community, etc. These two types of interactions collectively build the *social capital* of the community and are used to evaluate social trust.

**Temporal factor**: The decay of social trust with time is a fact of life in social networks. An interaction that has happened more recently may have more value than those that have happened some time back. Therefore, time is an important factor to capture the change in the behaviour of an individual.

**Context**: Suppose member X in the community trusts member Y for his or her recommendation of movies. This does not automatically mean that X trusts Y's views on a restaurant. Here, the movie represents the context of the trust between X and Y. It is important to distinguish between the different contexts of trust.

### *3.2 Social Capital Model*

Our social trust model STrust endeavours to capture all three elements described above, through *social capital*. Following (Brunie, 2009), we define the social capital of an online community *as the density of interactions that is beneficial to the members of the community*, *i.e., the positive interactions among the members in the community*. This is in contrast to existing trust literature (Maheswaran et al., 2007) (Hughes and Guttorp, 1994 ) where all interactions are treated equally, and passive interactions do not get much attention. In line with our definition of social trust, we separate social capital into two types: popularity and engagement. We further explain these using the graphical representation of interactions shown in Figure 2.

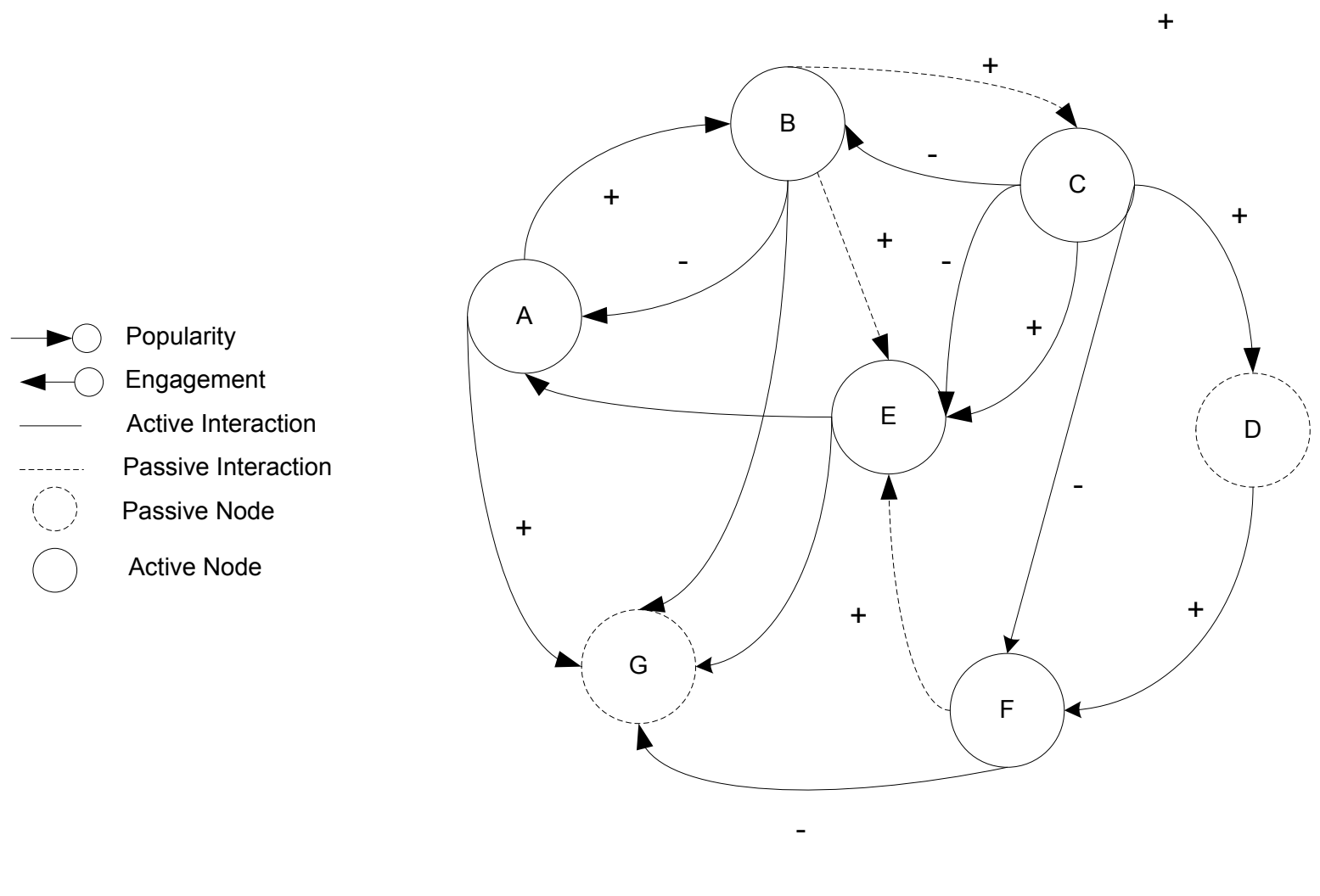

Figure 2. Network model of social capital

The nodes in the graph represent community members and the edges their interactions. The nodes could also be other entities in the community, e.g. activities or contents, which we refer to as passive nodes. Each arrow provides information towards popularity trust for one side (the sink or receiving end) and engagement trust on the other side (the source or initiating end). Let's consider node B as an example. It has four outgoing arrows and two incoming ones. The outgoing arrows support B's engagement trust, and the incoming ones support B's popularity trust. As another example, the arrow between A and B provides information towards the engagement trust of node A and the popularity trust of node B. Solid lines represent active interactions, and dotted lines represent passive interactions. Finally, the interactions between two nodes are either positive (represented as +) or negative (represented as -). Passive interactions are always considered positive.

### *3.3 Social Trust Model (STrust)*

We now describe how our social trust model *STrust* captures the three essential elements of social trust using the data captured by the social capital model described above. Our trust model contains the two types of trust derived from the two types of interactions defined earlier.

**Popularity Trust:** The popularity trust refers to the popularity of an individual member in the community. Metrics for the popularity trust can include the number of positive feedback received on the member's posts, the number of invitation requests for friendship received by the member, etc. Detailed description of the metrics is out of the scope of this paper. We model the popularity trust using beta family of probability distribution function (Josang and Ismail, 2002 ).

Let $U$ be the set representing the number of members in the community and $|PT_{ij}^{+}|$ the total number of positive interactions a member $u_i \in U$ has with the member $u_j \in U$. Similarly, the total number of negative interactions is represented as $|PT_{ij}^{-}|$. A member in the community may be involved in a number of activities related to a single context. A member may post a large number of messages in different contexts or within a single context. We need to consider this while evaluating the total number of positive (negative) interactions for a context ($x$) $|PT_{ij}^{x+}|$. $|PT_{ij}^{+}|$ and $|PT_{ij}^{-}|$ are computed as follows:

$$|PT_{ij}^{x+}| = \sum_{a=1}^{|A|} +1 \quad \text{and} \quad |PT_{ij}^{x-}| = \sum_{a=1}^{|A|} -1$$

where |X| represents the number of contexts, and |A| represents the number of activities in each context. The popularity trust (*PopTrust*) of a member $u_i \in U$ for a particular context ($x$) is then defined as:

$$PopTrust(u_i^x) = \frac{\sum_{j=1}^{|M|-1} \frac{|PT_{ij}^{x+}|+1}{|PT_{ij}^{x+}|+|PT_{ij}^{x-}|+2}}{|M|-1}$$

The aggregation over all contexts gives the popularity trust of the member in the community as follows:

$$PopTrust(u_i) = \frac{\sum_{x=1}^{|X|} PopTrust(u_i^x)}{|X|}$$

**Engagement Trust**: The engagement trust refers to the involvement of an individual member in the community. Example metrics for engagement trust include the number of positive feedback provided on the posts, the number of invitation requests sent to other members, etc. Engagements in a community can be of two types and hence the corresponding metrics: active and passive. In our model, we have considered both active and passive engagements. We define the engagement trust model in a similar way to the popularity trust model as follows.

$$EngTrust(u_i^x) = \frac{\sum_{j=1}^{|M|-1} \frac{|ET_{ij}^{x+}|+1}{|ET_{ij}^{x+}|+|ET_{ij}^{x-}|+2}}{|M|-1} \qquad EngTrust(u_i) = \frac{\sum_{x=1}^{|X|} EngTrust(u_i^x)}{|X|}$$

Where $|ET_{ij}^{x+}| = \sum_{a=1}^{|A|} +1$ , and $|ET_{ij}^{x-}| = \sum_{a=1}^{|A|} -1$

The social trust (*STrust*) of an individual member $u_i \in U$ in the community is then given by:

$$STrust(u_i) = \alpha.PopTrust(u_i) + (1-\alpha).EngTrust(u_i)$$

where $\alpha$ represents the value of a weight in the range of 0 to 1. If $\alpha$ =1, the social trust of an individual indicates how much other members in the community trust him or her. This is equivalent to the reputation of a member in the community. If $\alpha$ =0, the social trust represents how much a member trusts others in the community.

In ideal trust communities, all members in the society have almost the same high social trust. We can define the social trust of a community as:

$$STrust(c) = \frac{\sum_{i=1}^{|M|} SocialTrust(u_i)}{|M|}$$

An important question in this context is: what is an ideal value of social trust for a community to be called a *trust community*? Answering this question is challenging and interesting. In our work, we are bootstrapping the trust value for a community at 0.5. The social trust of the community must thus be greater than 0.5 for the community to be even considered a trust community. Any community that is considered a trust community is more likely to be a sustainable community.

## 4. Recommendation System

In the earlier section, we separated two types of interactions and derived three types of trust: popularity, engagement and social. One of the main reasons behind this separation is to use different trust values for different purposes. In this section, we propose a recommender system that exploits these different types of trust values to recommend different things to community members. The purpose of the recommendation system is to continuously build the social capital of the community by recommending new activities that lead to a new set of interactions among community members. In order to achieve this goal, the recommendation system needs to identify existing relationships among members in different contexts. We describe here the relationships in the community, how we evaluate them using trust values, and how we use trust values for recommendation.

**Member Relationships**: A relationship exists between two members in the community in a certain context of social life. For example, a member $u_i \in U$ has had positive experiences with his or her interactions with another member $u_j \in U$ about movies ($x$). The social trust between $u_i$ and $u_j$ for context $x$ is defined as follows:

$$STrust(u_i, u_j) = \alpha.PopTrust(u_{ij}, x) + (1-\alpha).EngTrust(u_{ij}, x)$$

Where $PopTrust(u_{ij}, x) = PopTrust(u_i, u_j, x) + PopTrust(u_j, u_i, x)$

$$PopTrust(u_i, u_j, x) = \frac{|PT_{ij}^{x+}| + 1}{|PT_{ij}^{x+}| + |PT_{ij}^{x-}| + 2}$$ and $$PopTrust(u_j, u_i, x) = \frac{|PT_{ji}^{x+}| + 1}{|PT_{ji}^{x+}| + |PT_{ji}^{x-}| + 2}$$

Similarly, $EngTrust(u_{ij}, x) = EngTrust(u_i, u_j, x) + EngTrust(u_j, u_i, x)$

$$EngTrust(u_i, u_j, x) = \frac{|ET_{ij}^{x+}|+1}{|ET_{ij}^{x+}|+|ET_{ij}^{x-}|+2} \quad \text{and} \quad EngTrust(u_j, u_i, x) = \frac{|ET_{ji}^{x+}|+1}{|ET_{ji}^{x+}|+|ET_{ji}^{x-}|+2}$$

where $PopTrust(u_i, u_j, x)$ represents the popularity trust of member $u_j$ towards member $u_i$ in context *x*.

We exploit these relationships to provide recommendations for members in the community. We discuss the recommendations that we have evaluated in this paper.

**Selecting leaders:** The online community provider may want to identify leaders in the community for different contexts/aspects of life. A leader is a member who is trustworthy in the community for providing useful posts/opinions/materials on a particular topic/context read/liked by many members. We can capture this in our model using the popularity component of social trust (i.e., $\alpha = 1$ ). This means a leader is the member who may have overall less social trust, but has a high popularity trust among members (e.g., has a high number of followers). A high level of engagement is not necessary for a leader. However, a certain level of engagement is needed to generate followers.

**Selecting mentors**: An online community provider may want to identify a likely mentor for a community member on a certain aspect. Recommendation of mentors fosters positive participation in the community and increases the social capital of the community. It is essential to have established trust relationships between a mentor and the member. In addition, it is important to determine that a mentor is a member who would like to engage in the community actively. This means the mentor must have a high level of engagement trust in the community (i.e., $\alpha \approx 0$ ).

In the next section, we verify the usefulness of the separation of different types of trust by analysing the recommendation of leaders and mentors on different types of network.

## 5. Analysis of the Model

Our purpose here is twofold: (a) show the utility of our model and (b) analyse the sustainability of social networks using the model. We carried out a few experiments using real datasets representing Facebook-like social networks. The data sets for the experiments were obtained from http://toreopsahl.com/datasets/ and represent interactions between students in an online community at the University of California, USA. These dataset have also been used to study network analysis of online community in (Panzarasa et al., 2009) and network clustering in (Opsahl and Panzarasa, 2009). The details of the datasets are described in Table 1.

The first dataset represents interactions between community members exchanging private messages, and the second dataset represents their interactions over a forum. Thus, in the first network, an interaction between two nodes is defined as a message exchanged between them, while, in the second network, a post in a forum is counted as an interaction. For our analysis, we do not consider the type of the post and simply take these two datasets to represent networks with different degree of granularity. The first dataset represents a network with more members but fewer interactions, while the second network has fewer members but a higher number of interactions. The purpose of choosing datasets with different characteristics is to observe the model in different settings. As the dataset does not contain information on whether an interaction is positive or negative, we assume it to be positive only. This information is enough for our study, as we intend to observe the trend of change in trust values rather than the computation of an exact figure.

| **Facebook like Social Network** | **Private Messages Data-Dataset I** | **Forum Interaction Data- Dataset II** |
|---|---|---|
| Total Number of members | 1899 | 899 |
| Total Number of interactions | 59835 | 1113924 |
| Number of unique interactions | 20296 | 142760 |

Table 1: Experiment Datasets

In the first part of our experiments, we aim to show the utility of the model. Here, we focus on the recommender system which recommends leaders and mentors in the community based on the trust model, as discussed in Section 4. Our experiments with the two datasets show how our model can effectively distinguish between these two types of community members. To analyse this, we first calculate the Engagement, Popularity and Social Trusts for each member in both networks. We then rank and filter members according to these three types of trust values. We specifically want to see if this distinction is a useful on, i.e., if members identified as popular (potential leaders) are different than members identified as engaged (potential mentors).

Figure 3 shows the overlap of members when selecting the top K leaders and mentors, where K ranges from 1 to 20. It is clear from the graph that the recommender recommends different sets of people for leaders and mentors. There is an overlap between these two sets, which varies from one data set to another. The trend clearly shows that the overlap in the data set I is lower than the one in data set II. Further analysis reveals it may not be necessary to separate popularity trust and engagement trust for a highly interactive network from the point of view of recommending leaders and mentors. However, this separation is useful for less interactive networks like the one represented by dataset I.

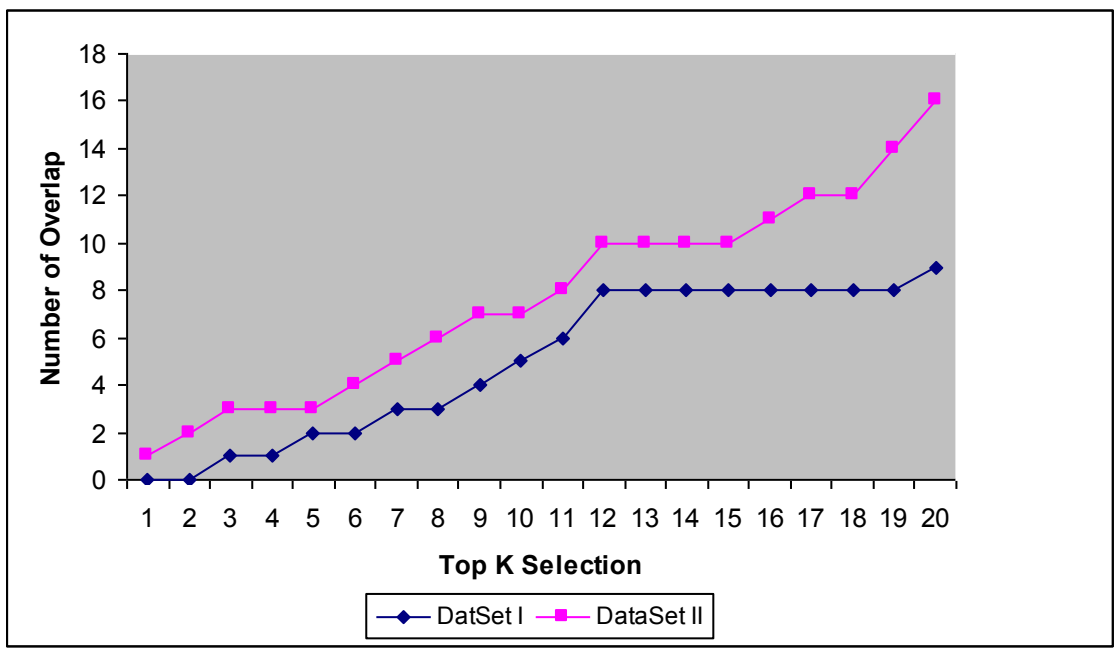

Figure 3. Overlap between top K selection of leaders and mentors

We now remove highly trusted people in the community and identify the total number of unique members (non-overlapping members) that are eliminated. A high number of unique members justifies the separation of different types of trust and thus validates the model. Elimination is done as a percentage of the total members in the network, at intervals of 5, 10 and 15%. We compare the distinctness of members eliminated when they are ranked according to social, engagement and popularity trusts. A member node is unique if it is eliminated from the ranked list of only one type of trust. For example, member 9 is a unique member in the elimination of "popularity trust and engagement trust" if it appears in the engagement trust ranked list (i.e., a candidate for removal with respect to the engagement trust), but does not appear in the ranked list for the popularity trust, or vice versa. As shown in Figure 4 (a), there are 32 unique members between engagement and popularity trust when 5% of the top ranked members are eliminated from both lists. That is, a recommender recommends 32 members *either* as leaders or mentors. Therefore, our results show that, in real datasets, we can distinguish between different types of members.  Figure 4 presents the statistics of the comparisons for the three types of trust. The results also validate the insights gained from the top K selection, as the Dataset II has fewer unique eliminations in comparison to Dataset I (Figure 4 (b)).

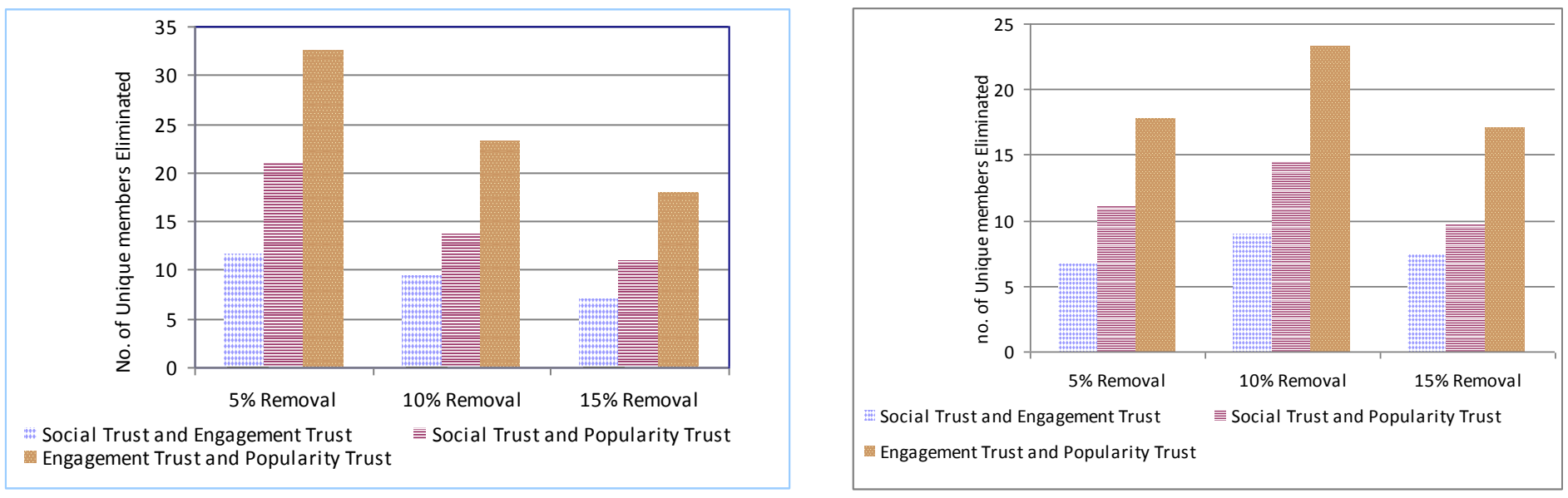

(a) (b)

Figure 4. Comparison of the unique eliminations in different settings, (a). Data Set I and (b). Data Set II

We next study the sustainability issue. Specifically, we show the significance that a critical mass of trustworthy members has in assuring the sustainability of any online network. First, we calculate the three different types of trust as before. We then rank the nodes according to their values. Highly ranked members are eliminated from the network to observe how the community social trust is affected by their absence. A node in the network can be visualised through its interconnections with other nodes in the community. The same node might have different views for its engagement and popularity in the network. As an example, we choose 'Node 9' from the first dataset. This node ranked first as a highly engaged node in the community, while its position in terms of popularity in the same network was beyond 50. Figures 5(a) and 5 (b) offer visualisation of this node from dataset I. The direction of the arrows in the engagement visualisation is outwards, whereas, in the popularity visualisation, they concentrate towards the centre.

The first thing we observe is the influence that the removal of members with high social trust has on the community social trust. Figure 6 shows the community social trust and social capital for both the datasets, before the removal were made. The charts in figure 7 (a) show the decreasing trend of the community social trust when 5, 10 and 15% of the members with high social trust are removed from the community. It is important to note that the social trust model is bootstrapped at 0.5 (with the presence of 1 in numerator and 2 in denominator), and all interactions are assumed to be positive; so the social trust values always appear above 0.5.

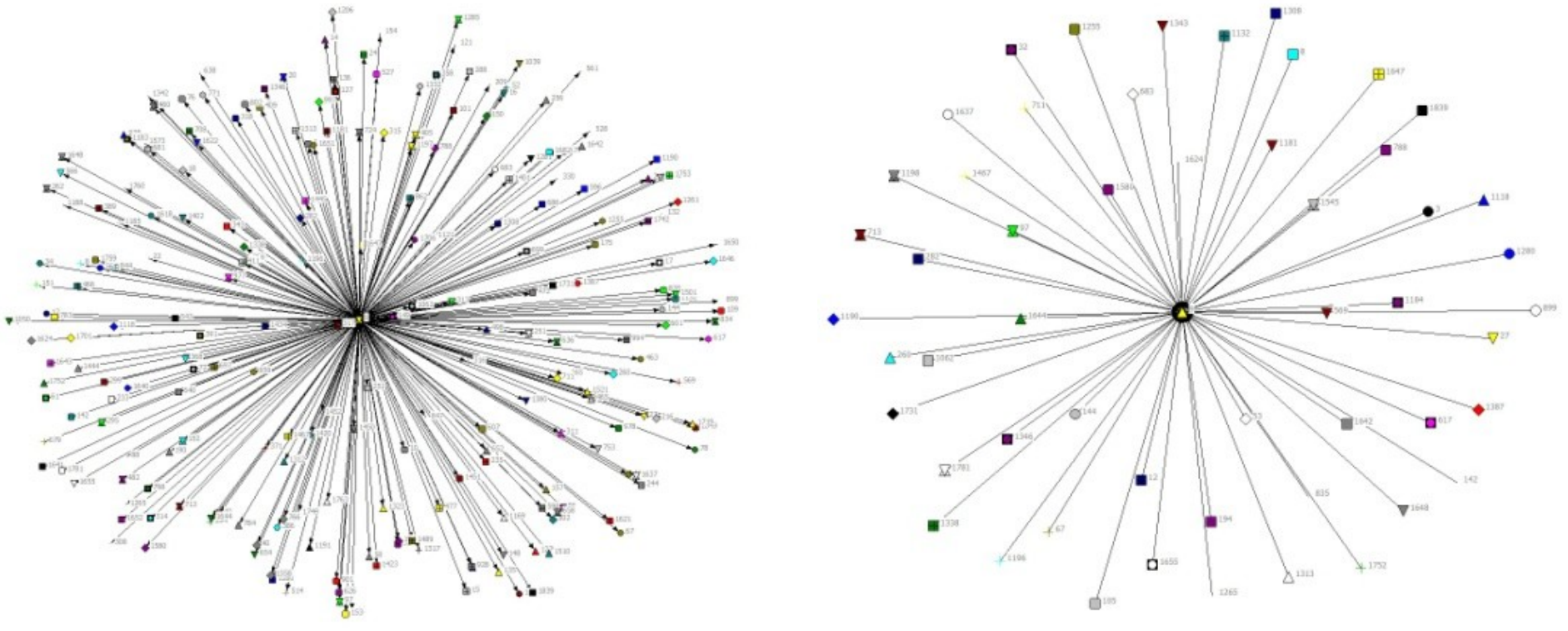

**(a)** **(b)**

Figure 5. (a) Engagement Visualisation, (b) Popularity Visualisation of Node 9 Data Set I

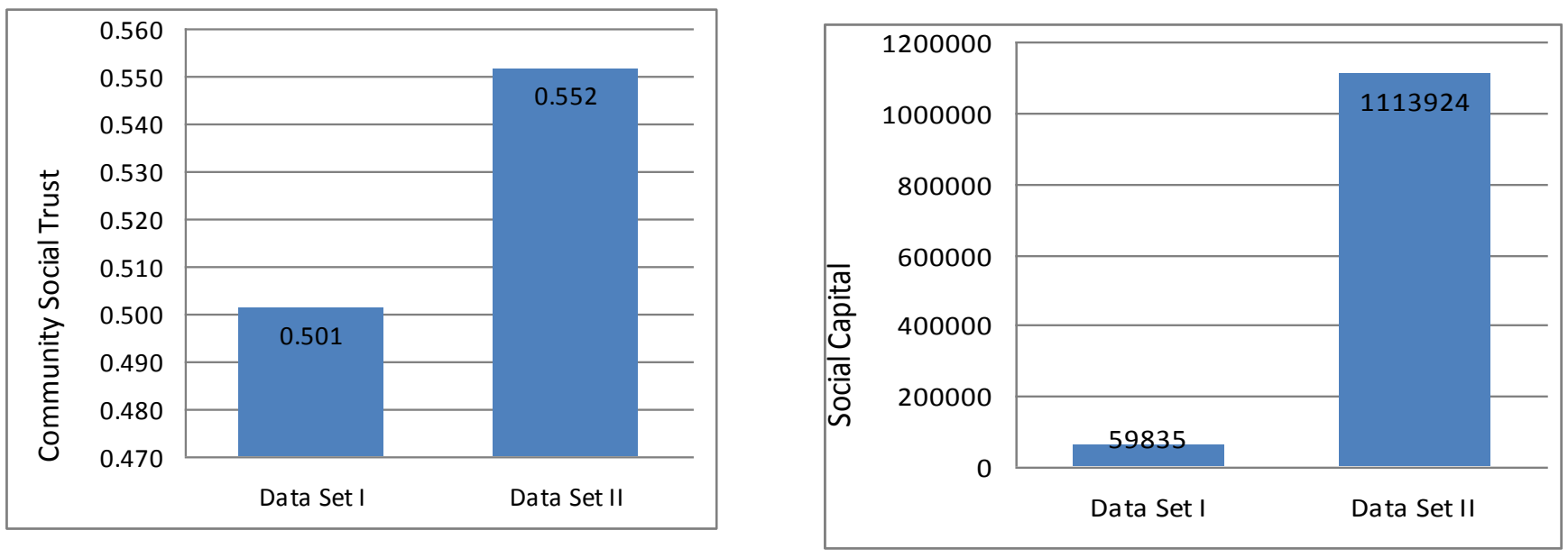

Figure 6. Community Social Trust and Social Capital Values for the full networks

This can be explained by the lower social capital (interactions) of dataset I comparing to data set II. In a community with low social capital, removal of active members has a high impact on the community social trust. This gives an interesting prediction on the sustainability of a community. Specifically, what type of community might last longer? To investigate this, we study the trend of unique interactions in the community. For both datasets and for all removals, we compute the percentage difference between the total interactions and the total unique interactions. (If member X interacts with member Y, it is a unique interaction, no matter how many times they interact with each other. If they interact with each other many times, there will be many interactions, but only one unique interaction.) Figure 7(b) shows the result for this computation.

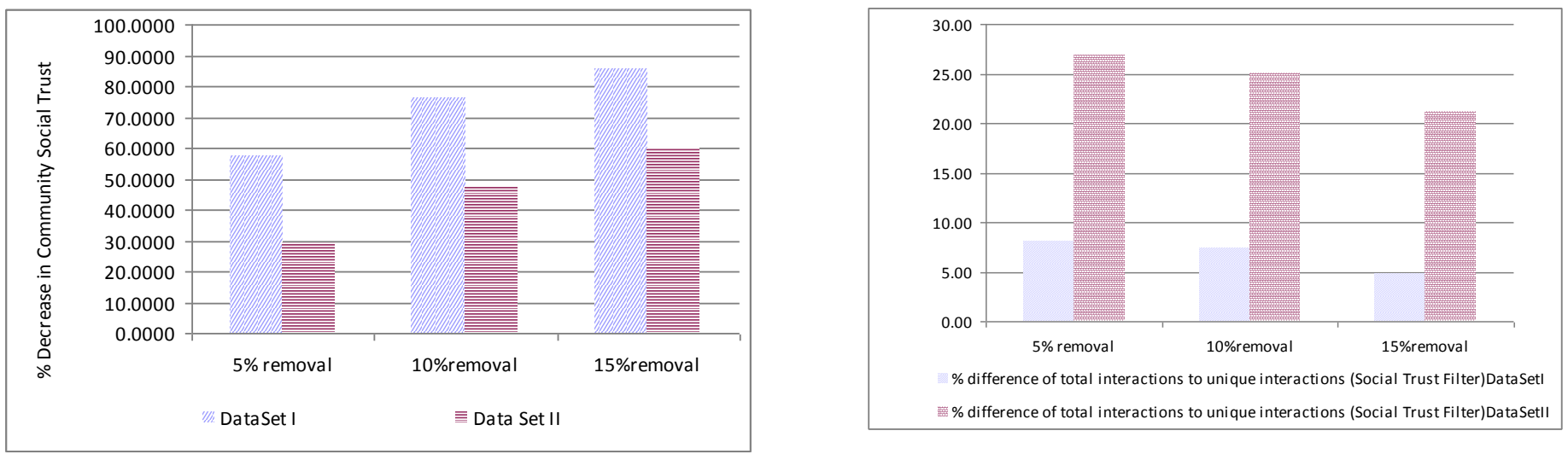

(a)

(b)

Figure 7. (a) Percentage decrease in Community Social Trust with increasing removal, and (b) Percentage difference between total interactions and unique interactions

We can see that, with increasing removal, the percentage difference between the total number of interactions and total number of unique interactions is narrowing. Eventually the gap would continue to narrow until it reaches a value nearing zero, indicating that there are not many multiple interactions in the community; at this point, the community may have a problem of sustainability, as all members are interacting with each other at most once. The results here show that, for dataset I, at 15% removal, the percentage difference between total interactions and total unique interactions is 4.81%, while it is 21.81% for data set II. This means that if a community is more interactive (like that represented by data set II) and has a higher social capital, it is more sustainable and thus better equipped to withstand loss of interactive members as well.

In both cases, community social trust has decreased with the increasing proportion of elimination from the community. Interestingly, elimination in dataset I has greater impact than in dataset II. In case of 15% removal, dataset I suffered an 84% decrease in community social trust, whereas, for the same setting, dataset II suffered a 60% loss (note that the percentage drop is calculated by removing the bootstrapping value 0.5). As dataset II is more interactive as compared to dataset I, the impact of removing nodes with higher social trust is less severe on the community social trust than in data set II. Furthermore, data set I sees a higher reduction in the total number of interactions as compared to dataset II, as shown in Figure 8.

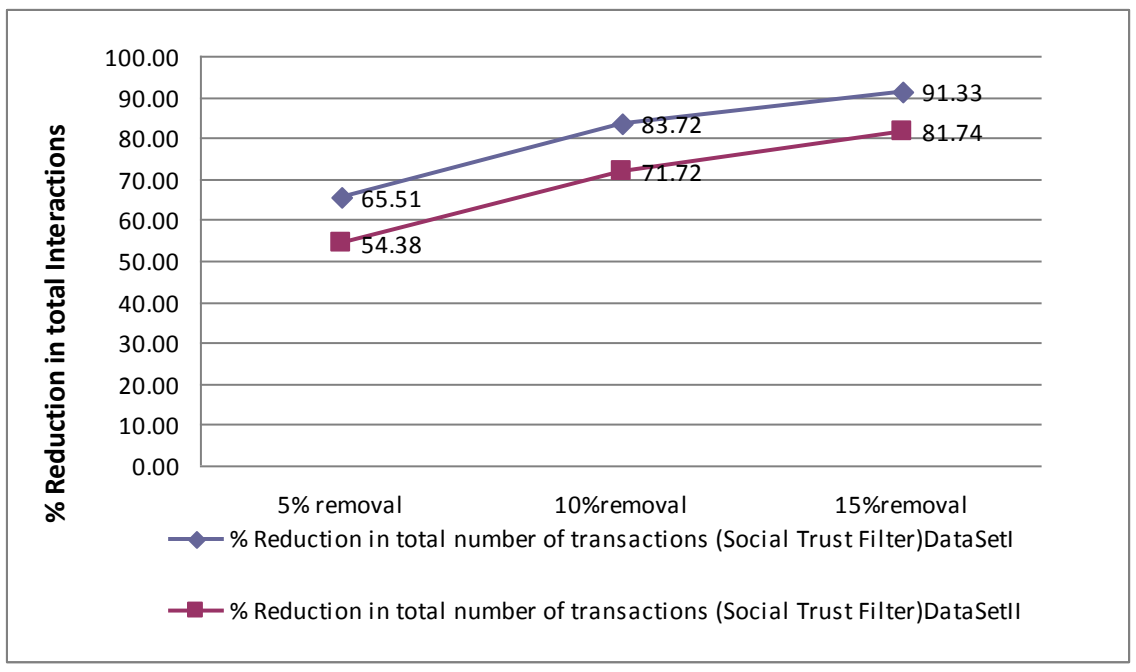

Figure 8. Percentage of interactions reduced with increasing removal

## 6. Related Work

This section focuses on the literature that covers two aspects of social networks discussed in this paper: (a) recommender systems, and (b) viability and sustainability of social networks.

Recommendation systems can be broadly categorised as: (i) content-based, (ii) collaborative filtering-based, (iii) hybrid and (iv) trust-based. Content-based approaches produce recommendations based on the similarity between items consumed by the members. The root of the content-based methods lies in Information Retrieval (IR). Examples of such recommenders include InfoFinder (Krulwich and Burkey, 1996) and NewsWeeder (Lang, 1995). Collaborative filtering approaches recommend the items chosen by users with similar tastes/preferences (Sarwar et al., 2001). In contrast to similarity between items in content-based recommenders, this approach uses similarity between users. Example systems taking this approach include GroupLens (Resnick et al., 1994) and Ringo (Shardanand and Maes, 1995). In order to overcome the shortcomings of pure content and collaborative based recommendation systems, hybrid approaches have been used, where items are recommended to users when they score highly against their own profile, and when they are rated highly by a user with a similar profile. An example of such systems includes Fab (Balabanovi and Shoham, 1997). In recent times, trust based recommendation systems have gained popularity. They usually construct a trust network where nodes are users, and edges represent the trust placed on them. The goal of a trust-based recommendation system is to generate personalised recommendations by aggregating the opinions of other users in the trust network.

Walter *et al.* (2009) propose a model for computing indirect trust between two agents which are not neighbours based on the direct trust between agents that are neighbours. The direct trust refers to the trust derived from interactions between two agents, whereas the indirect trust refers to the trust derived using transitive or propagation properties. This model makes use of the social network structure for computing trust and the computed indirect trust is then used to generate recommendations. The recommendation algorithm combines content-based recommendation with trust between the nodes to generate recommendations. Similarly, Zarghami *et al.* (2009) address the sparsity and scalability of collaborative filtering by introducing the concept of T-index, similar to H-index used to measure the science research output. Recommendation techniques that analyse trust networks were found to provide very accurate and highly personalised results. Hang *et al.* (2010) use a graph-based approach to recommend a node in a social network using similarity in trust networks. Massa *et al.* (2007) propose a trust-based recommendation system where it is possible to search for trustable users by exploiting trust propagation over the trust network. Andersen *et al*. (2008) explore an axiomatic approach for trust-based recommendation and propose several recommendation models, some of which are incentive compatible (i.e., malicious members cannot entice other members to provide false/misleading trust information and trust links because it is always in the interest of the member to provide factual information). Our approach in this paper exploits the social trust model to recommend different things and roles to different people as discussed in earlier sections.

Another important aspect discussed in this paper is the sustainability of social networks. When you search the phrase “social networks come and go” in Google and follow a few top ranked links, you will understand what we meant. You will find a number of articles where authors questioned the sustainability of social networks. The primary point of these articles is that social networking sites are great when they serve your purpose and you have friends on them. Yeomans & Warner (2011) studied the use social networks for business sustainability. However, there is limited research on the sustainability of social networks themselves. Buttler (2001) studied the role of size and communication activity in sustainable online social structures. He presented a resource-based theory of sustainable social structures. The focus of this study was on the resource constraints, i.e., the ability of the community to attract and retain members when the size of the membership is greater than the available resources (more specifically, communication resources). Though Buttler’s underlying assumption in this paper on scarcity of network bandwidth for higher level of communication activity in the community is no longer valid, the model is equally applicable in situations where there is a scarcity of resources to be consumed by members in the community. Simpson (2005) studied the sustainability of community informatics and identified that social capital matters for effective widespread uptake and the sustainability of community informatics initiatives. In line with this study, we analyse the sustainability of social networks by using a social trust model derived from social capital.

Network analysis is also a topic of interest in computing. Behavioural analysis and the evolution of a web based social network was done by (Kumar et al., 2006) by selecting two real datasets from Yahoo and Flickr. An interesting finding in this work is the pattern in which the network evolves. With both the datasets, the authors have shown that the networks first see a rapid growth followed by decline and then a slow but steady growth. In an analysis of a huge social network reported in (Ahn et al., 2007), the authors have analysed degree distribution, clustering property, degree correlation and evolution over time for three online social network services: Cyworld, MySpace and Orkut, each having over 10 million users at the time of study. Interestingly their research shows that heterogeneity in a network leads to multi-scaling behaviour in degree distribution. Cyworld data was shown to have a multi-scaling degree distribution as compared to simple scaling behaviours exhibited by MySpace and Orkut data. Adamic *et al.* in (Buyukkokten et al., 2003) have studied attributes that contribute to friendship formation in social network. Choosing a Stanford online social network called Club Nexus, they have measured network parameters like clustering, strength of weak ties etc. and shown how similarity between members decay with increase in the network separation. Measures of centrality and connectedness in scientific network was done by (Newman, 2001). Computer database of scientific papers in physics, biomedical research and computer science was used to construct the collaboration network and study collaboration patterns. Differences in collaboration patterns between subjects studied have been reported.

## 7. Conclusions and Future Work

In this paper, we have briefly presented our social trust model, which decouples two types of trust (engagement and popularity trusts), and its accompanying recommender system, capable of recommending leaders and mentors as distinctive entities. We analysed two types of social networks using our model. The objectives of our analysis were: (a) to validate the separation of engagement and popularity trusts, and (b) to utilise our model to study the sustainability issue. For the first objective, we looked at the overlap (or, conversely, the uniqueness) of leaders and mentors as would be recommended by our system. We observed that a significant number of unique members are recommended in these two different categories. This result validates the separation of popularity trust from engagement trust. Our analysis also shows that the number of unique members is lower in the highly interactive community then in the community with lower number of interactions.

We next studied the sustainability of social networks by exploiting the social trust model. Our idea was to observe the decrease in social capital when removing highly trusted members. Our hypothesis is that we need a minimum number of highly trusted members to sustain a community. We computed popularity and engagement trusts for each individual member. We then removed highly trusted members at intervals of 5% of the total population. Our analysis shows that the social capital of the networks decreased by more than 50% when 5% of the highly trusted members and their interactions were eliminated from the community. This went up to 80% when 15% were eliminated. Our analysis has also provided the insight that the percentage difference between the total interactions and unique interactions is higher in the highly interactive community as compared to the community with the lower number of interactions.

Our analysis has a number of limitations that we plan to address in future work. First, all interactions in our datasets were positive. Further analysis on datasets with both positive and negative interactions is required to understand the implication of negative interactions on social capital. Second, the effect of temporal decay or aging of interactions on social capital still needs to be studied (our data had no time information). Finally, we want to study social networks that were sustained and some that are dead. We believe these further analyses will shed some light on what percentage of members in the social networks need to be positively/negatively active to keep them sustained.